\pdfoutput=1
\documentclass[aps,prl,reprint,amssymb,amsmath,superscriptaddress]{revtex4-1}

\usepackage{graphicx}%
\usepackage{dcolumn}%
\usepackage{bm}%
\usepackage{siunitx}%
\usepackage{bbm} %
\usepackage[pdftex]{hyperref}%
\usepackage{wasysym}

\usepackage{xr} %
\newcommand{\Airy}[1]{\text{Airy}_{#1}}
\newcommand{\Airyproc}[2]{\mathcal{A}_{#1}\left(#2\right)}

\newcommand{\ave}[1]{\left<#1\right>}

\newcommand{\cum}[2]{\left<#1^{#2}\right>_{\rm c}}

\newcommand{\chiprbl}[1]{{\tilde{\chi}\left({#1}\tau\right)}}

\newcommand{\ind}{\mathbbm{1}}
\newcommand{\tod}{\stackrel{\mathrm{d}}{\to}}
\newcommand{\eqd}{\stackrel{\mathrm{d}}{=}}
\newcommand{\simeqd}{\stackrel{\mathrm{d}}{\simeq}}
\newcommand{\expct}[1]{\langle #1 \rangle}

\newcommand{\Sk}[1]{\mathrm{Sk}[#1]}  %
\newcommand{\Ku}[1]{\mathrm{Ku}[#1]}  %

\newcommand{\smallO}[1]{{\scriptstyle \mathcal{O} }\left(#1\right)}

\makeatletter
\@ifclasswith{revtex4-1}{linenumbers}{\newcommand{\equationlinenobegin}{\begin{linenomath}}}{\newcommand{\equationlinenobegin}{}}
\@ifclasswith{revtex4-1}{linenumbers}{\newcommand{\equationlinenoend}{\end{linenomath}}}{\newcommand{\equationlinenoend}{}}
\makeatother

\begin{document}

\title{Kardar-Parisi-Zhang Interfaces with Curved Initial Shapes and Variational Formula}%
\author{Yohsuke T. Fukai}
\email{ysk@yfukai.net}
\affiliation{%
	RIKEN Center for Biosystems Dynamics Research
}%
\affiliation{%
	Department of Physics, the University of Tokyo
}%
\author{Kazumasa A. Takeuchi}%
\email{kat@kaztake.org}
\affiliation{%
	Department of Physics, the University of Tokyo
}%

\date{\today}

\begin{abstract}
We study fluctuations of interfaces in the Kardar-Parisi-Zhang (KPZ) universality class with curved initial conditions. 
By simulations of a cluster growth model and experiments of liquid-crystal turbulence, we determine the universal scaling functions that describe the height distribution and the spatial correlation of the interfaces growing outward from a ring.
The scaling functions, controlled by a single dimensionless time parameter, show crossover from the statistical properties of the flat interfaces to those of the circular interfaces.
Moreover, employing the KPZ variational formula to describe the case of the ring initial condition, we find that the formula, which we numerically evaluate, reproduces the numerical and experimental results precisely without adjustable parameters.
This demonstrates that precise numerical evaluation of the variational formula is possible at all, 
and underlines the practical importance of the formula, which is able to predict the one-point distribution of KPZ interfaces for general initial conditions.
\end{abstract}

\maketitle

Efforts on universal behavior associated with scale invariance,
 which have established important concepts
 such as the renormalization group and the universality class,
 now shed light on novel aspects of nonequilibrium fluctuations.
In this respect, the Kardar-Parisi-Zhang (KPZ) universality class
\cite{kardar_1986,barabasi_1995,[{For recent reviews, see, e.g., }][]kriecherbauer_2010,*corwin_2012,*quastel_2015,*halpin-healy_2015,*[][{.}]sasamoto_2016,[{For lecture notes, see: }][]takeuchi_2018} plays a distinguished role, because of the existence of exact solutions and experimental realizations.
The KPZ class is also known to arise in a variety of problems: besides growing interfaces and directed polymers as originally proposed \cite{kardar_1986},
 it also turned out to be relevant for stochastic particle transport,
 quantum integrable systems \cite{kriecherbauer_2010,takeuchi_2018},
 and fluctuating hydrodynamics \cite{spohn_2016}, to name but a few.

In the following, let us focus on the one-dimensional case, 
 for which exact studies have been developed, and consider growing interfaces 
 described by the height $h(x,t)$
 at position $x\in\mathbb{R}$ and time $t\in\mathbb{R}$.
The KPZ class describes scale-invariant fluctuations of growing interfaces
in the long-time limit, 
in general situations without particular symmetries and conservation laws.
The hallmark of the KPZ class is the scaling laws for the fluctuation amplitude $\sim{}t^\beta$ and the correlation length $\sim{}t^{1/z}$, with universal exponents $\beta$ and $z$ that take the values $\beta=1/3$ and $z=3/2$ for the one-dimensional case \cite{kardar_1986,barabasi_1995,takeuchi_2018}.
The height $h(x,t)$ is then generally written, for large $t$, as 
\equationlinenobegin\begin{equation} \label{eq:h_asymptotic}
 h(x,t) \simeq v_\infty t + \left(\Gamma t\right)^{1/3} \chi(X,t)
\end{equation}\equationlinenoend
 where $\chi(X,t)$ is a stochastic variable,
 $X:=x/\xi(t)$ denotes the coordinate rescaled by
 the correlation length $\xi(t):=\frac{2}{A}\left(\Gamma{}t\right)^{2/3}$,
 and $v_\infty,\Gamma,A$ are system-dependent parameters.
The variable $\chi(X,t)$ is expected to be universal,
 in the sense that its statistical properties
 do not depend on microscopic details of the systems.
The scaling exponents of the KPZ class have been found in various experimental systems \cite{takeuchi_2014}, including colonies of living cells \cite{wakita_1997,huergo_2010,*huergo_2011,*huergo_2012}, combusting paper \cite{maunuksela_1997}, and liquid-crystal turbulence \cite{takeuchi_2018,takeuchi_2010,*takeuchi_2011,takeuchi_2012,fukai_2017}.

Recently, remarkable developments triggered by exact studies
 \cite{kriecherbauer_2010,takeuchi_2018}
 have unveiled novel aspects on the KPZ class.
A particularly important outcome is the geometry dependence,
 which we describe below.
If an interface grows on top of a flat substrate,
 as usually assumed in simulations,
 the interface roughens but maintains the globally flat profile.
In contrast, if an interface in a plane starts to grow from a point nucleus,
 say, at $x=0$, it takes a circular shape with a growing radius.
Although this interface becomes flatter and flatter
 as the radius %
 increases,
 statistical properties of $\chi(X,t)$ remain distinct from the flat case.
Specifically, $\chi(X,t)$ has different asymptotic behavior as follows
\equationlinenobegin\begin{equation}
 \chi(X,t) \tod \begin{cases} 
\mathcal{A}_1(X), & \text{(flat)} \\
\mathcal{A}_2(X) - X^2, & \text{(circular)}
\end{cases}  \label{eq:Airy}
\end{equation}\equationlinenoend
 where $\tod$ denotes convergence in distribution
 ($\eqd$ and $\simeqd$ will be used analogously).
$\mathcal{A}_1(X)$ and $\mathcal{A}_2(X)$ are called
 the Airy$_1$ \cite{sasamoto_2005,borodin_2007a} and Airy$_2$ \cite{prahofer_2002} processes,
 respectively, and well studied analytically \cite{quastel_2014}. 
Due to their translational invariance, as long as one-point properties are concerned, 
 $\mathcal{A}_i(X)$ can be replaced by a single stochastic variable $\chi_i$.
Remarkably, the one-point distribution of $\chi_1$ and $\chi_2$
 was shown \cite{baik_1999,johansson_2000,Baik.Rains-MSRIP2001,prahofer_2000a,*prahofer_2000} to coincide respectively with the GOE and GUE
 Tracy-Widom distribution \cite{Tracy.Widom-CMP1994,*Tracy.Widom-CMP1996}, 
 known from random matrix theory \cite{anderson_2010},
 which describes the distribution of the largest eigenvalue of random matrices
 in the Gaussian orthogonal and unitary ensembles (GOE and GUE).
This geometry dependence,
 as well as the emergence of the Tracy-Widom distribution,
 turned out to be experimentally relevant too,
 as shown by experiments on liquid-crystal turbulence
 \cite{takeuchi_2010,*takeuchi_2011, takeuchi_2012,takeuchi_2018}.
Correlation properties were also shown to be different
 between the flat and circular cases, even though the scaling exponents
 $\beta$ and $z$ take the same values.
On the basis of those results, 
one may state that the flat and circular interfaces constitute different \textit{universality subclasses} 
within the single KPZ class, characterized by different yet universal distribution and correlation properties.

Those universality subclasses have been, however, mostly studied
 for a few ``canonical'' cases including the flat and circular ones.
A natural and important question is then what happens for more general initial conditions.
Theoretically, the \textit{KPZ fixed-point variational formula} 
\cite{corwin_2015,corwin_2016b,dauvergne_2019,quastel_2014,quastel_2016} 
can be used to predict the asymptotic properties 
 of $\chi(X,t\to\infty)$ for general initial conditions. 
On the other hand, experimental and numerical studies have focused
 on finite-time behavior emerging from intermediate initial conditions.
For example, the present authors \cite{fukai_2017} studied
 growth from a ring of finite radius $R_0$, which then produces
 two curved interfaces, one growing outward and the other one inward.
Focusing on the ingrowing interfaces, 
 we found that finite-time properties of $\chi(X,t)$ for different $R_0$
 are controlled solely by the rescaled time $\tau:=v_\infty{}t/R_0$,
 as follows: statistical properties of $\chi(X,t)$
 agree with those for the \textit{flat} subclass initially ($\tau\ll{}1$),
 until the interfaces nearly collapse at $\tau\approx{}1$
 and therefore do not behave as KPZ anymore.
Analogous behavior was also observed numerically by Carrasco and Oliveira
 \cite{carrasco_2018a}, who used lattice models
 with system size set to decrease in time
 (mimicking the shrinking circumference of the ingrowing interfaces).
The case of enlarging substrates,
 which would correspond to the outgrowing case,
 has also been studied and crossover from the flat to circular subclasses 
 was suggested in this case \cite{carrasco_2014,carrasco_2018a,carrasco_2018},
 which is also expected to be described by $\tau$.
However, it remains unclear how universal such finite-time behavior is,
 why $\tau$ is the right parameter to describe it, and above all, 
 how such crossover can be described theoretically.

Those problems are addressed and answered in this Letter.
We study outgrowing interfaces from ring initial conditions
 both numerically and experimentally,
 using an off-lattice version of the Eden model \cite{takeuchi_2012a} 
 and the liquid-crystal turbulence
 \cite{takeuchi_2010,takeuchi_2012,fukai_2017,takeuchi_2018}.
Scaling functions for the flat-to-circular crossover are determined, and
 shown to be the same for both of the studied systems.
Moreover, we describe this crossover theoretically,
 by adapting the variational formula \cite{corwin_2015,corwin_2016b,dauvergne_2019,quastel_2014,quastel_2016} 
 for curved initial conditions.
The formula is numerically evaluated and shown to reproduce our numerical and experimental results quantitatively, without adjustable parameters.
This also implies that the flat-to-circular crossover is indeed universal
 and, furthermore, should generally appear for \textit{any} curved interfaces
 with locally parabolic initial conditions.

\begin{figure}
\includegraphics[width=\linewidth]{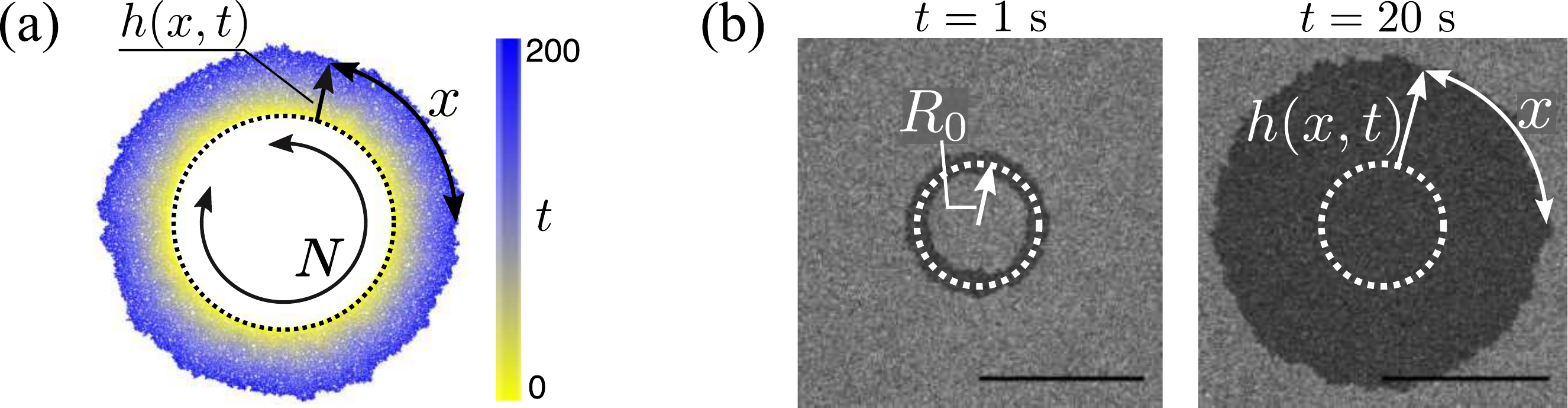}
\caption{\label{fig:schematic}
Typical snapshots from the Eden simulations and the liquid-crystal experiments.
(a) An Eden interface growing outward from a ring
 with $N=1000$ (dotted line). 
Time is indicated by the color. 
(b) A DSM2 cluster (black) growing from a ring with $R_0=\SI{366}{\um}$
 (dotted lines). 
The elapsed time after shooting laser is indicated above each image. 
The scale bar corresponds to $\SI{1}{\mm}$. 
}
\end{figure}

We first study the off-lattice Eden model \cite{takeuchi_2012a},
 in which a cluster of round particles (with unit diameter)
 grows by stochastic addition of new particles.
The initial condition is set to be a ring of $N$ particles
 [Fig.~\ref{fig:schematic}(a)].
The evolution rule is as follows (see Ref.~\cite{takeuchi_2012a} for details):
 at each time step, we randomly choose a particle at the interface,
 attempt to put a new particle next to it in a random direction
 and do so if there is no overlapping particle.
Time is then increased by $1/\text{(the number of the interfacial particles)}$
 whether the new particle was put or not.
Particles that cannot contribute further growth were checked and removed from the list of the interfacial particles every time unit.
To characterize the height fluctuations,
 we measure the local radius increment $R(\theta,t)$,
 which is the radial distance between the initial ring and the interface
 at each angular position $\theta$ [Fig.~\ref{fig:schematic}(a)].
Thanks to the rotational symmetry, we have
\equationlinenobegin\begin{equation}
 R(\theta,t) \eqd h(0,t) \simeq v_\infty t + (\Gamma t)^{1/3}\chi(0,t),
\end{equation}\equationlinenoend
 but statistical precision can be improved by averaging over $\theta$.
In our simulations, we varied the initial size $N$ from 100 to 40000
 and obtained 4320 to 14400 realizations for each case
 (summarized in Table~\ref{S-tab:simulation_params_eden} \cite{supplementary}).
For comparison, we also simulated flat interfaces,
 for which the initial condition was a line formed by $75000$ particles
 and the periodic boundary condition in the spanwise direction was used,
 and obtained 14400 realizations.

To characterize statistical properties
 of the stochastic variable $\chi(X,t)$, 
 we first estimated the non-universal parameters $v_\infty,\Gamma$ and $A$,
 from the data for the flat interfaces.  
$v_\infty$ and $\Gamma$ were obtained by the standard procedure \cite{takeuchi_2018}, specifically by
 using $\partial_t\ave{h}\simeq{}v_\infty+\mathrm{const.}\times{}t^{-2/3}$
 and $\cum{h}{2}/\left(t^{2/3}\cum{\chi_1}{2}\right)\simeq\Gamma^{2/3}$,
 where $\cum{\cdots}{k}$ denotes the $k$th-order cumulant
 and here we used the fact that
 the asymptotic fluctuations of the flat interfaces
 are given by the GOE Tracy-Widom distribution.
We obtained $v_\infty=0.51370(5)$ and $\Gamma=0.980(3)$.
The parameter $A$ was obtained by $A=\sqrt{2\Gamma/v_\infty}$,
 the relationship valid for isotropic growth \cite{takeuchi_2012}.

\begin{figure}
\includegraphics[width=\linewidth]{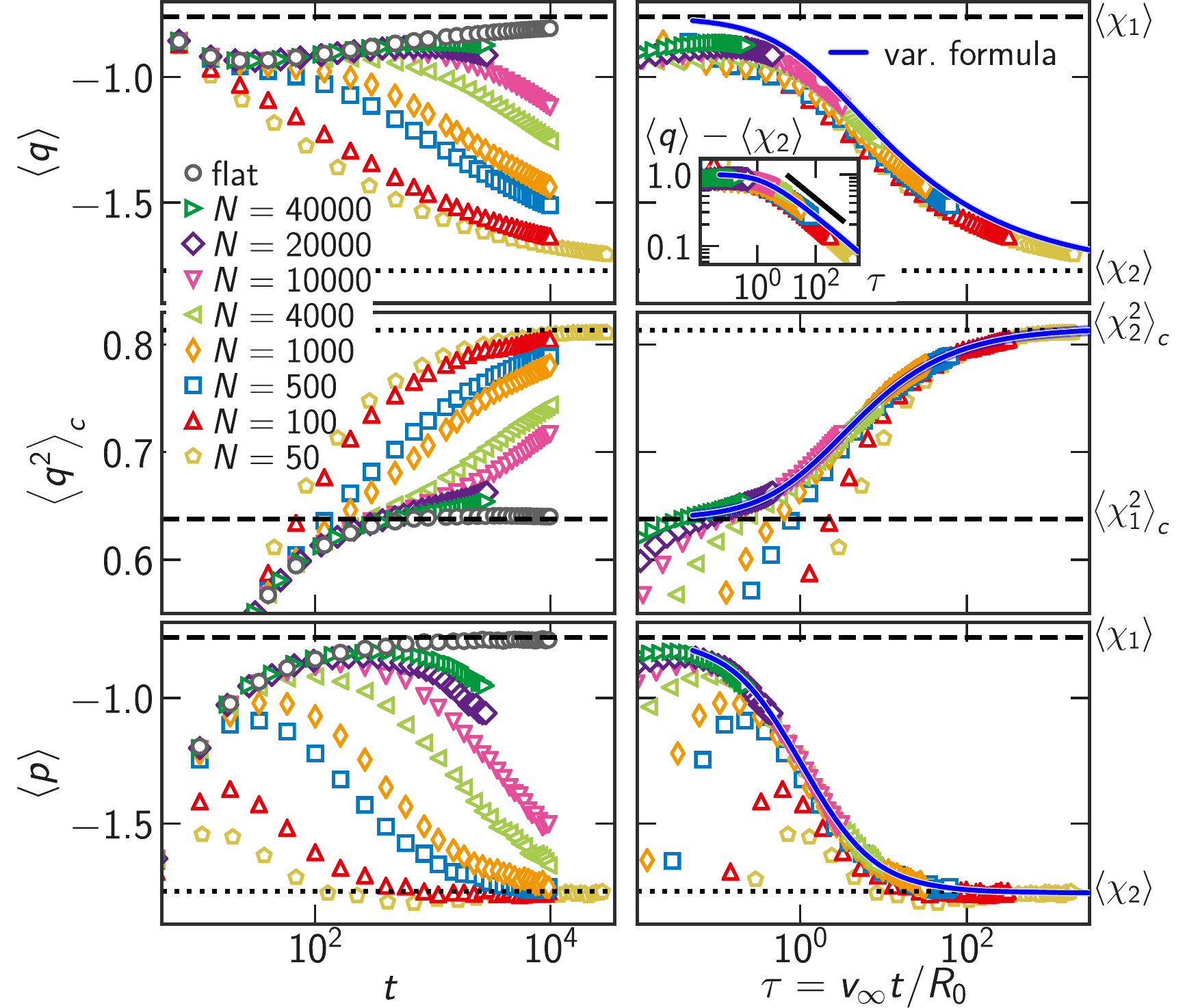}
\caption{\label{fig:eden}
The mean and variance of the rescaled height,
 $\expct{q(\theta,t)}$ and $\cum{q(\theta,t)}{2}$,
 and the rescaled mean velocity $\expct{p(\theta,t)}$
 for the Eden model in the outgrowing case. 
The data are shown against the raw time $t$ (left)
 and the rescaled time $\tau=v_\infty{}t/R_0$ (right). 
The theoretical curves evaluated numerically from the variational formula 
 for the outgrowing interfaces (=var., blue solid line)
 are shown in the right panels.
The values of $\chi_1$ and $\chi_2$ are shown by the dashed and dotted lines,
 respectively. 
The inset of the right-top figure shows the difference between the data and the excepted long-time limit value, $\ave{\chi_2}$. 
The black solid line indicates slope $-1/3$.
}
\end{figure}

With those parameter values, we define the rescaled height
\equationlinenobegin\begin{align}\label{eq:q_def} 
 \begin{split}
 q(\theta,t):=\frac{R(\theta,t)-v_\infty t}{\left(\Gamma t\right)^{1/3}} \simeqd \chi(0,t)
 \end{split}
\end{align}\equationlinenoend
 and measure its mean and variance as functions of time,
 for different initial particle number $N$
 (Fig.~\ref{fig:eden} left).
Figure~\ref{fig:eden} also shows the rescaled mean velocity \cite{fukai_2017,takeuchi_2018} 
\equationlinenobegin\begin{align}
		\left<p(\theta,t)\right>&:=\left<\frac{3t^{2/3}}{\Gamma^{1/3}}\left[\partial_t R(\theta,t)-v_\infty\right]\right> \notag \\
		&\simeq\left<\chi(0,t)\right> +3 t \partial_t \left<\chi(0,t)\right>,\label{eq:p_asymptotic}  
\end{align}\equationlinenoend
 which asymptotically goes to $\left<\chi(0,t)\right>$
 if $\left<\chi(0,t)\right>$ converges sufficiently fast.
For the flat case (gray circles), 
 $\expct{q}\to\expct{\chi_1},\expct{p}\to\expct{\chi_1}$
 and $\cum{q}{2}\to\cum{\chi_1}{2}$ as expected.
In the case of the ring initial conditions, for large $N$
 the data first behave similarly to the flat case, then deviate and approach
 the values for the circular subclass, $\ave{\chi_2}$ and $\cum{\chi_2}{2}$ 
 \footnote{Though the value of $\ave{q}$ does not fully converge to $\ave{\chi_2}$ even at the largest $\tau$ we reached, 
 	the difference seems to converge to zero with a power law with an exponent close to $-1/3$ (inset of Fig.~\ref{fig:eden}).  
 	This suggests convergence of $\ave{q}$ to $\ave{\chi_2}$ in the limit of $t\to\infty$.}.  
This crossover takes place earlier for smaller $N$.
Indeed, when the data are plotted
 against the rescaled time $\tau=v_\infty t/R_0$ ($R_0=N/2\pi$),
 all data collapse onto a single curve
 except for the non-universal short-time regime (Fig.~\ref{fig:eden} right).
This suggests that the distribution of $\chi(0,t)$ for different $R_0$,
 denoted by $\chi(0,t;R_0)$, is described
 by a single stochastic variable $\chi_\mathrm{c}(0,\tau)$,
 parametrized by $\tau$, as follows:
\equationlinenobegin\begin{align}
 \chi(0,t;R_0) \tod \chi_\mathrm{c}(0,\tau),  \qquad (R_0, t\to \infty)
 \label{eq:crossover_fluctuation}
\end{align}\equationlinenoend
 where the double limit is taken with fixed $\tau=v_\infty{}t/R_0$.
Then the flat-to-circular crossover we found indicates
 $\chi_\mathrm{c}(0,\tau)\tod\chi_1$ for $\tau\to0$ and
 $\chi_\mathrm{c}(0,\tau)\tod\chi_2$ for $\tau\to\infty$.
The skewness $\Sk{R(\theta,t)}\allowbreak:=\allowbreak\cum{R}{3}/\cum{R}{2}^{3/2}\allowbreak\to\allowbreak\Sk{\chi_\mathrm{c}(0,\tau)}$
 and the kurtosis $\Ku{R(\theta,t)}\allowbreak:=\allowbreak\cum{R}{4}/\cum{R}{2}^2\allowbreak\to\allowbreak\Ku{\chi_\mathrm{c}(0,\tau)}$ 
 show consistent behavior
 (Fig.~\ref{S-fig:eden_skew_kurt} \cite{supplementary}).

We also study this crossover in the spatial correlation.
In the case of the point initial condition,
 suppose $\theta=0$ corresponds to $x=0$,
 then using $R(\theta,t)=\sqrt{h(x,t)^2+x^2}\simeq{}h+\tfrac{x^2}{2h}$
 and Eq.~\eqref{eq:Airy},
 we can show $q(\theta,t)\tod\mathcal{A}_2(X)$.
Therefore, the rescaled spatial covariance
$C_\mathrm{s}(\Delta{}X,t):=\expct{q(\theta+\Delta\theta,t)q(\theta,t)} - \expct{q(\theta,t)}^2$
 with $\Delta{}X:=\expct{R(\theta,t)}\Delta\theta/\xi(t)$
 can be directly compared with the covariance
 of the Airy$_1$ and Airy$_2$ processes.
Our numerical results for the ring initial conditions
 (Fig.~\ref{S-fig:spatial_corr} filled symbols) indeed show crossover
 from the $\Airy{1}$ covariance ($\tau\ll{}1$)
 to the $\Airy{2}$ covariance ($\tau\gg{}1$),
 consistently to the results on the one-point distribution.

To test universality of our finding, 
 in particular the function forms
 of $\expct{\chi_\mathrm{c}(0,\tau)}$ and $\cum{\chi_\mathrm{c}(0,\tau)}{2}$,
 we conducted experiments on liquid-crystal turbulence \cite{takeuchi_2010,takeuchi_2011,takeuchi_2012,takeuchi_2018,fukai_2017}. 
As in the previous studies,
 we applied an AC voltage (here, $\SI{22}{\V}$ at $\SI{300}{\Hz}$)
 to nematic liquid crystal filling a thin gap between transparent electrodes, 
 and observed growth of a turbulent state
 called the dynamic scattering mode 2 (DSM2),
 expanding in a metastable turbulent state, DSM1
 (see Supplemental Text \cite{supplementary} for detailed methods). 
DSM2 was generated by emitting a few ultraviolet laser pulses
\cite{takeuchi_2018}.
Using the holographic technique we previously adopted
 for the DSM2 growth experiments \cite{fukai_2017},
 we formed the laser intensity profile
 in the shape of a ring of a given radius $R_0$,
 which sets the initial condition of the DSM2 interface [Fig.~\ref{fig:schematic}(b)].
We also generated circular interfaces with a point initial condition,
 and flat interfaces with a linear initial condition.
We obtained 941 to 1936 realizations for each case
 (Table~\ref{S-tab:nonuniversal_params_lc} \cite{supplementary}),
 recorded by a charge-coupled device camera.
The radius $R(\theta,t)$ of the DSM2 interfaces
 (or the height $h(x,t)$ for the flat case) was determined
 from each image, with the time $t$ defined
 as the elapsed time after shooting the laser pulses.
Then the non-universal parameters $v_\infty,\Gamma,A$ were
 evaluated in the same way as for the Eden model,
 here for the flat and point initial conditions
 (Table~\ref{S-tab:nonuniversal_params_lc} \cite{supplementary}).
Although the values of $v_\infty,\Gamma,A$ are expected to be independent
 of the initial condition, in practice one needs to evaluate
 for each set of experiments, because of unavoidable slight changes
 in experimental conditions \cite{takeuchi_2012}.
For the ring initial conditions, however, the parameter values
 could not be obtained in the same way because of the time dependence
 (i.e., crossover) of $\chi(X,t)$.
We therefore used the values obtained from the flat case
 for the outgrowing cases, unless otherwise stipulated.
Possible shifts in the parameter values were taken into account
 in the uncertainty estimates for the outgrowing cases,
 evaluated from the differences in the parameter values
 between the flat and circular cases.

\begin{figure}
	\includegraphics[width=\linewidth]{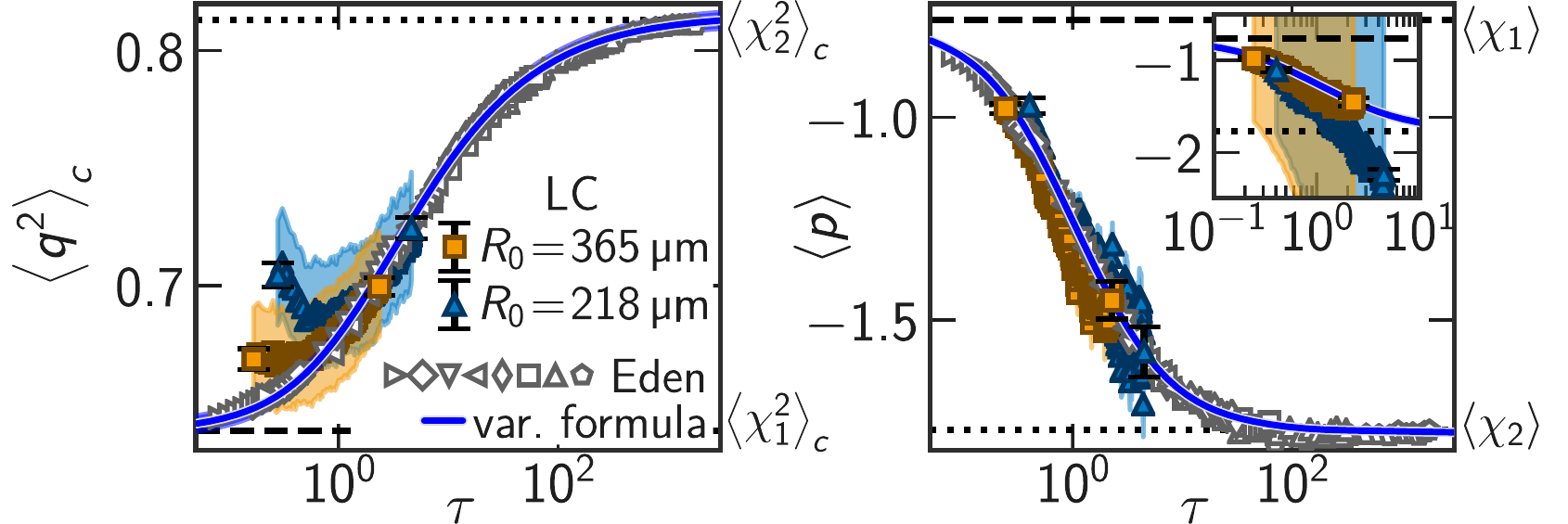}
	\caption{\label{fig:out_cum}
Comparison of the results from the experiments (color filled symbols),
 the Eden simulations (gray open symbols),
 and the variational formula (=var., blue solid line),
 for the outgrowing interfaces.
The variance of the rescaled height, $\cum{q(\theta,t)}{2}$,
 and the rescaled mean velocity $\expct{p(\theta,t)}$ are shown
 in the left and right panels, respectively, against $\tau=v_\infty{}t/R_0$.
For the numerical results, data with $t>10^3$ are shown by the same
symbols as those in Fig.~\ref{fig:eden}. 
For the experimental results, statistical errors are indicated
 by the error bars on the first and last data points, and
 uncertainty associated with the parameter estimation is shown
 by the shaded areas. 
The values for $\chi_1$ (flat) and $\chi_2$ (circular)
 are shown by the dashed and dotted lines, respectively.
The inset of the right panel shows the experimental results
 obtained with $v_\infty$ from the flat case,
 while it was adjusted in the main panel
 to fit the Eden data at the largest $t$ (see text).
	}
\end{figure}

Now we compare the experimental results with those for the Eden model.
Figure~\ref{fig:out_cum} left panel shows the variance of the rescaled height,
 $\cum{q(\theta,t)}{2}$, against $\tau=v_\infty{}t/R_0$,
 which overlaps on the Eden data
 within statistical errors and parameter uncertainty
 (error bars and shades, respectively)
 apart from the non-universal short-time behavior.
For the rescaled mean velocity $\expct{p(\theta,t)}$ (right panel),
 the uncertainty of $v_\infty$ was too large
 to make a meaningful comparison (inset).
However, if we instead choose the value of $v_\infty$ in such a way that
 $\expct{p(\theta,t)}$ at the largest $t$ falls
 onto the curve for the Eden model (obtained values of $v_\infty$ are given
 in Table~\ref{S-tab:nonuniversal_params_lc}),
 $\expct{p(\theta,t)}$ overlaps for \textit{all} $t$ (main panel).
Those results of $\cum{q(\theta,t)}{2}$ and $\expct{p(\theta,t)}$
 suggest universality of the one-point distribution
 of $\chi_\mathrm{c}(0,\tau)$.
Moreover, the spatial covariance $C_\mathrm{s}(\Delta{}X,t)$
 is also found to overlap with the results of the Eden model
 if the value of $\tau$ is close enough
 (Fig.~\ref{S-fig:spatial_corr}).
This suggests that
 not only the one-point distribution of $\chi_\mathrm{c}(0,\tau)$
 but the spatial covariance of $\chi_\mathrm{c}(X,\tau)$ is also universal.

So far we have characterized the flat-to-circular crossover
 and found it to be controlled by a single parameter $\tau=v_\infty{}t/R_0$, 
 but why so and how can this crossover be theoretically described? 
To answer these questions, we employ the variational formula
\cite{corwin_2015,corwin_2016b,dauvergne_2019,quastel_2014,quastel_2016} 
 and apply it to a general, curved initial condition.
 
The variational formula describes the height $h(x,t)$
for a general initial condition $h(x,0)=:h_0(x)$ 
as follows 
\equationlinenobegin\begin{equation}
 h(x,t) \simeqd \sup_{y\in\mathbb{R}}\left[h_\mathrm{circ}(x,t;y) + h_0(y) \right], \label{eq:var_formula_h}
\end{equation}\equationlinenoend
where $h_\mathrm{circ}(x,t;y)$ denotes the height for the point initial condition 
nucleating at position $y$, 
growing with the same realization of noise for different $y$ \cite{corwin_2015}.
Intuitively, this means that the initial condition $h(x,0)$ can be regarded
 as a collection of point sources and $h(x,t)$ is then given
 by the envelope of the circular interfaces from those point sources,
 a bit analogously to Huygens' principle
\footnote{
Though nonlinear equations generally
 do not admit the principle of superposition, 
 the KPZ equation can be mapped to a linear equation
 by the Cole-Hopf transformation and this leads to the variational formula \eqref{eq:var_formula_h}
 \cite{quastel_2014}. 
}.
The formula \eqref{eq:var_formula_h} involves a mathematical object called the Airy sheet \cite{corwin_2015,dauvergne_2019},
 but if the interest is only in the one-point distribution,
 it can be simply expressed by the Airy$_2$ process, as follows \cite{quastel_2014,corwin_2016b}:
\equationlinenobegin\begin{equation}
 \chi(X,t) \simeqd 
 \sup_{Y\in\mathbb{R}}\left[\mathcal{A}_2(X-Y) - (X-Y)^2 + \frac{h_0(\xi(t)Y)}{(\Gamma t)^{1/3}} \right]. \label{eq:var_formula_general}
\end{equation}\equationlinenoend

We use Eq.~\eqref{eq:var_formula_general} and consider a class of curved initial conditions in the following form
\equationlinenobegin\begin{equation}
h_0(x)=R_0 g\left(\frac{x}{R_0}\right) \label{eq:curved_h_0}
\end{equation}\equationlinenoend
 where $g(w)$ is a locally parabolic function, i.e.,
 $g(w)=-c_2w^2+\smallO{w^2}$ for small $|w|$.
Substituting Eq.~\eqref{eq:curved_h_0} into Eq.~\eqref{eq:var_formula_general},
 taking the limit $R_0,t\to\infty$ with fixed $\tau=v_\infty{}t/R_0$,
 and setting $x=0$ yields 
\equationlinenobegin\begin{equation}
 \chi(0,t) \tod 
\sup_{Y\in\mathbb{R}}\left[\Airyproc{2}{Y} - \left(1+c \tau \right) Y^2 \right] =: \chiprbl{c} \label{eq:var_formula_parabola_general}
\end{equation}\equationlinenoend
 with $c:=(4c_2\Gamma)/(A^2v_\infty)$.
This shows that the asymptotic height distribution 
is parameterized only by $c\tau$, 
and only the local functional form of $g\left(w\right)$ at small $\left|w\right|$ is relevant.
The characteristic time is $\tau=1/c$ and therefore $t=A^2R_0/4c_2\Gamma$,
 and this is the time at which the initial height difference $\left|h_0(0)-h_0(\xi\left(t\right))\right|$ becomes comparable to the fluctuation amplitude, $\left(\Gamma{}t\right)^{1/3}$.
For isotropic growth, the relationship $A=\sqrt{2\Gamma/v_\infty}$
 \cite{takeuchi_2012} further yields $c=2c_2$. 

For the ring initial conditions, $g(w)$ is given by 
 $g(w)=\sigma\left(\sqrt{1-w^2}\ind_{\left|w\right|<1}-1\right)$ 
 with $\sigma=+1$ ($-1$) for the outgrowing (ingrowing) case.
Then we obtain $\chi(0,t)\simeqd\chiprbl{\sigma}$,
 which we have expressed by $\chi_\mathrm{c}(0,\tau)$
 for the outgrowing case $\sigma=+1$ [Eq.~\eqref{eq:crossover_fluctuation}].
Note that, mathematically, it is known that
 $\tilde{\chi}(0)=\sup_{Y\in\mathbb{R}}(\Airyproc{2}{Y}-Y^2)\eqd\chi_1$,
 i.e., GOE Tracy-Widom distribution \cite{johansson_2003,Corwin.etal-CMP2013}.
In the other limit $\tau\to\infty$, clearly,
 $\tilde{\chi}(\tau)\to\Airyproc{2}{0}\eqd\chi_2$,
 i.e., GUE Tracy-Widom distribution.
Therefore, $\chi_\mathrm{c}(0,\tau)=\tilde{\chi}(\tau)$ indeed
 has the expected limits on both sides of the flat-to-circular crossover.

To compare the variational formula with the experimental and numerical data
 for finite $\tau$, 
 we employ a Monte Carlo method to evaluate
 Eq.~\eqref{eq:var_formula_parabola_general}.
The Airy$_2$ process $\mathcal{A}_2(Y)$ is in fact known to be equivalent
 to the largest eigenvalue of large GUE random matrices
 undergoing Dyson's Brownian motion \cite{johansson_2003,quastel_2014}.
We therefore implement Dyson's Brownian motion numerically,
in the form of the Ornstein-Uhlenbeck process of %
Hermitian random matrices 
 and obtained approximated realizations of $\mathcal{A}_2(Y)$ (see Supplemental Text \cite{supplementary} for details).
Then we evaluated the supremum of Eq.~\eqref{eq:var_formula_parabola_general},
 interpolating the values of $\mathcal{A}_2(Y)$ between the discrete steps
 by using the Brownian bridge \cite{supplementary}.
The results for the outgrowing case ($\sigma=+1$) are shown
 in Figs.~\ref{fig:eden} and \ref{fig:out_cum}, where the data of
 the mean $\ave{q}$, variance $\cum{q}{2}$, and the rescaled mean velocity $\expct{p}$ are
 compared with the corresponding expressions of $\tilde{\chi}(\tau)$,
 specifically, $\ave{\tilde{\chi}(\tau)}$, $\cum{\tilde{\chi}(\tau)}{2}$ [Eq.~\eqref{eq:q_def}], 
 and $\ave{\tilde{\chi}(\tau)}+3\tau\partial_\tau\ave{\tilde{\chi}(\tau)}$
 [Eq.~\eqref{eq:p_asymptotic}], respectively.
The results of the variational formula precisely agree,
 without any adjustable parameter, with the numerical and experimental data.
We also inspected the ingrowing case $\sigma=-1$
 and confirmed the validity of the variational formula (Fig.~\ref{S-fig:in_cum}).
The agreement was also underpinned for the skewness and kurtosis
 (Fig.~\ref{S-fig:outin_skew_kurt}).

In summary, we found KPZ crossover functions that govern height fluctuations
 of interfaces growing outward from ring initial conditions,
 parameterized only by the rescaled time $\tau=v_\infty{}t/R_0$,
 and evidenced their universality both experimentally and numerically.
We then presented a theoretical description of this crossover,
 on the basis of the KPZ variational formula
 for general curved initial conditions.
We numerically evaluated the formula and found remarkable agreement
 with the experimental and numerical data.
Our results constitute the first example where the KPZ variational formula was
 successfully used to describe experimental observations,
 showing the ability of this formula to explain, or even predict,
 real data from general initial conditions.
We hope our work will trigger further studies to elucidate geometry-dependent universality of the KPZ class and beyond.

\begin{acknowledgments}
We thank P. Le Doussal for useful discussions on the variational formula,
 and F. Bornemann for the theoretical curves
 of the $\text{Airy}_1$ and $\text{Airy}_2$ covariance
 \cite{Bornemann-MC2010}.
We thank Supercomputer Center of the Institute for Solid State Physics (the University of Tokyo) and 
Meiji Institute for Advanced Study of Mathematical Sciences (Meiji University)
 for computational facilities. 
We acknowledge financial support by KAKENHI
 from Japan Society for the Promotion of Science
 (Grant Nos. JP25103004, JP16H04033, JP19H05800, JP19H05144, JP17J05559),
 by Yamada Science Foundation,
 and by the National Science Foundation
 (Grant No. NSF PHY11-25915). 
\end{acknowledgments}

\bibliography{citations}
\end{document}

\title{Supplemental Material for \\
	Kardar-Parisi-Zhang Interfaces with Curved Initial Shapes and Variational Formula}%
\author{Yohsuke T. Fukai}
\email{ysk@yfukai.net}
\affiliation{%
	RIKEN Center for Biosystems Dynamics Research
}%
\affiliation{%
	Department of Physics, the University of Tokyo
}%
\author{Kazumasa A. Takeuchi}%
\email{kat@kaztake.org}
\affiliation{%
	Department of Physics, the University of Tokyo
}%

\date{\today}

\maketitle

\onecolumngrid

\section{Supplemental Text 1: Experimental Methods}

Similarly to the past studies
 \cite{takeuchi_2010,takeuchi_2011,takeuchi_2012,takeuchi_2018,fukai_2017}, 
 we prepared an electroconvection cell by assembling two glass plates with transparent electrodes, sandwiching $\SI{12}{\um}$-thickness spacers. 
A liquid-crystal sample, \textit{N}-(4-Methoxybenzylidene)-4-butylaniline 
 doped with $0.01\mathrm{wt.\%}$ of tetra-\textit{n}-butylammonium bromide, 
 was filled in a $\SI{1.5}{\cm}\times\SI{1.5}{\cm}$ region enclosed by the
 spacers.
The homeotropic alignment was achieved by spin-coating \textit{N},\textit{N}-dimethyl-\textit{N}-octadecyl-3-aminopropyltrimethoxysilyl chloride 
on the electrodes.
The cell was contained in a temperature controller, whose temperature was maintained at $\SI{25}{\celsius}$, 
with the temporal fluctuation in the order of $\pm\SI{0.01}{\celsius}$.
Under this temperature,
 the electroconvection was observed by applying an AC voltage to the cell.
The cutoff frequency, which separates the conductive and dielectric regimes of the electroconvection \cite{p.g.degennes_1995}, was roughly $\SI{1.7e3}{\Hz}$.
At the frequency we used in the main experiments, $\SI{300}{\Hz}$,
 compact DSM2 clusters grew at the amplitude $\gtrsim \SI{17.5}{V}$.
The voltage amplitude for the main experiments, $\SI{22}{\V}$,
 was chosen to be sufficiently higher than that threshold.

To generate a growing DSM2 interface,
 we first applied an AC voltage
 of amplitude $\SI{22}{\V}$ and frequency $\SI{300}{\Hz}$ to the cell
 and the system was set entirely in the metastable DSM1 state.
After $\SI{5}{s}$, we emitted three ultraviolet laser pulses to the cell
 (New Wave Research MiniLaseII, wavelength $\SI{355}{nm}$, pulse width $4$-$\SI{6}{ns}$, repetition frequency $\SI{20}{Hz}$, energy $\lessapprox\SI{0.4}{\milli\J}$ after attenuation)
 to nucleate DSM2 \cite{takeuchi_2018}.
The laser pulse was reflected
 by a spatial light modulator (Hamamatsu Photonics, LCOS-SLM X10468-05)
 and a hologram of the given shape was made,
 by using the experimental setup reported in Ref.~\cite{fukai_2017}.
For the linear initial condition,
 the line length was approximately $\SI{8}{\mm}$.
Then the growing DSM2 interface was recorded by a charge-coupled device camera
 for a given time 
(for the flat interfaces, the region of width $\approx\SI{4}{\mm}$ near the center of the line was observed). 
Then we turned off the applied voltage, waited for $\SI{30}{s}$,
 and started the next run.

To obtain the radius $R(\theta,t)$ of the DSM2 interfaces
 (or the height $h(x,t)$ for the flat case), 
 we binarized each image by thresholding.
For the ring initial conditions, first we determined the center of the ring
 by using the ensemble average of the image intensity fields
 taken at $t=\SI{1}{\s}$, and used it to define the radius $R(\theta,t)$.

\section{Supplemental Text 2: Evaluation of the variational formula}

\subsection{Detailed description of the numerical method}

\begin{figure}[H]
	\includegraphics[width=\linewidth]{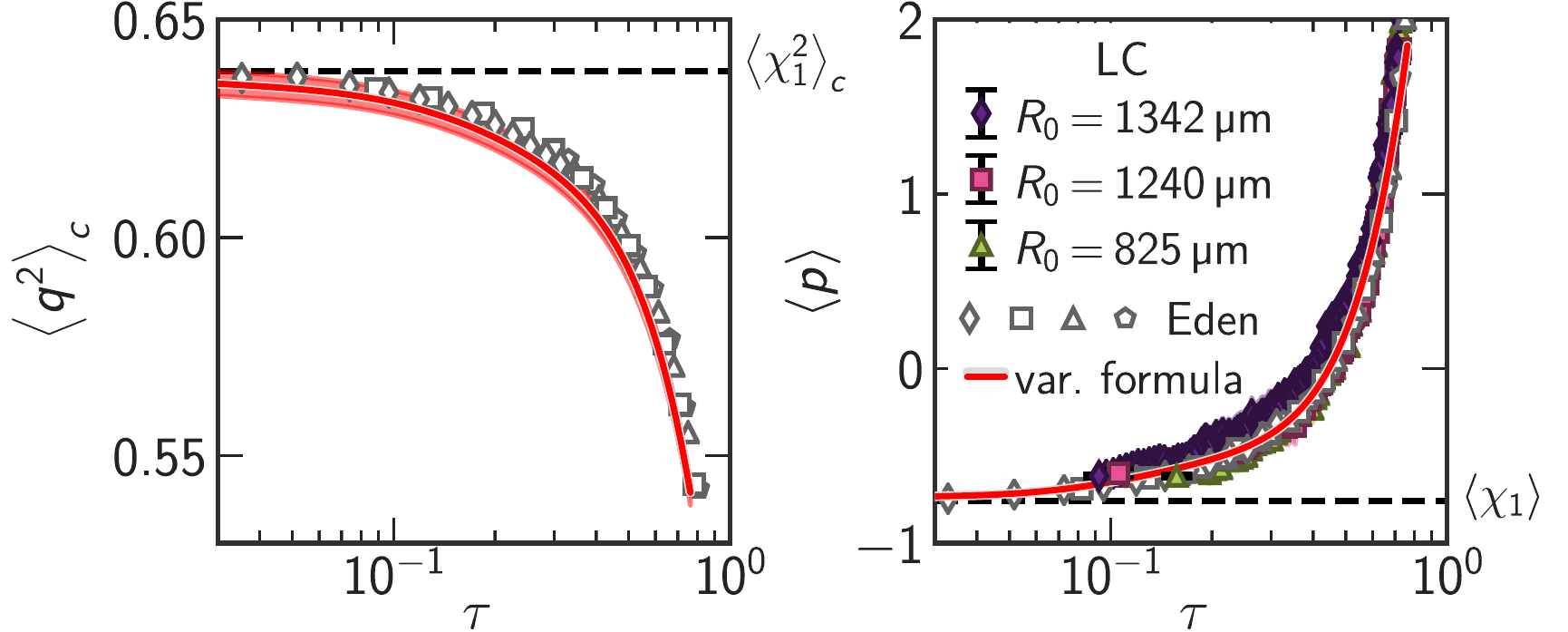}
	\caption{\label{fig:in_cum}
		The ingrowing counterpart of Fig.~\ref{fig:out_cum}.
		Results from the liquid-crystal experiments
		(color filled symbols; data adopted from Ref.~\cite{fukai_2017})
		and those from Eden simulations
		[gray open symbols;
		$N=100000(\Diamond),40000(\square),20000(\triangle),10000(\pentagon)$]
		are compared with the curves obtained from the variational formula
		(=var., red solid line).
		The shaded area behind the variational formula curves indicates uncertainty of the Monte-Carlo evaluation.
		See Table~\ref{tab:simulation_params_eden} for the number of realizations
		for the simulations.
		The experimental data for the variance (left) is omitted
		because of large finite-time effect \cite{fukai_2017}.
		The values for $\chi_1$ (flat) are shown by the dashed lines.
	}
\end{figure}

\begin{figure}[H]
	\includegraphics[width=\linewidth]{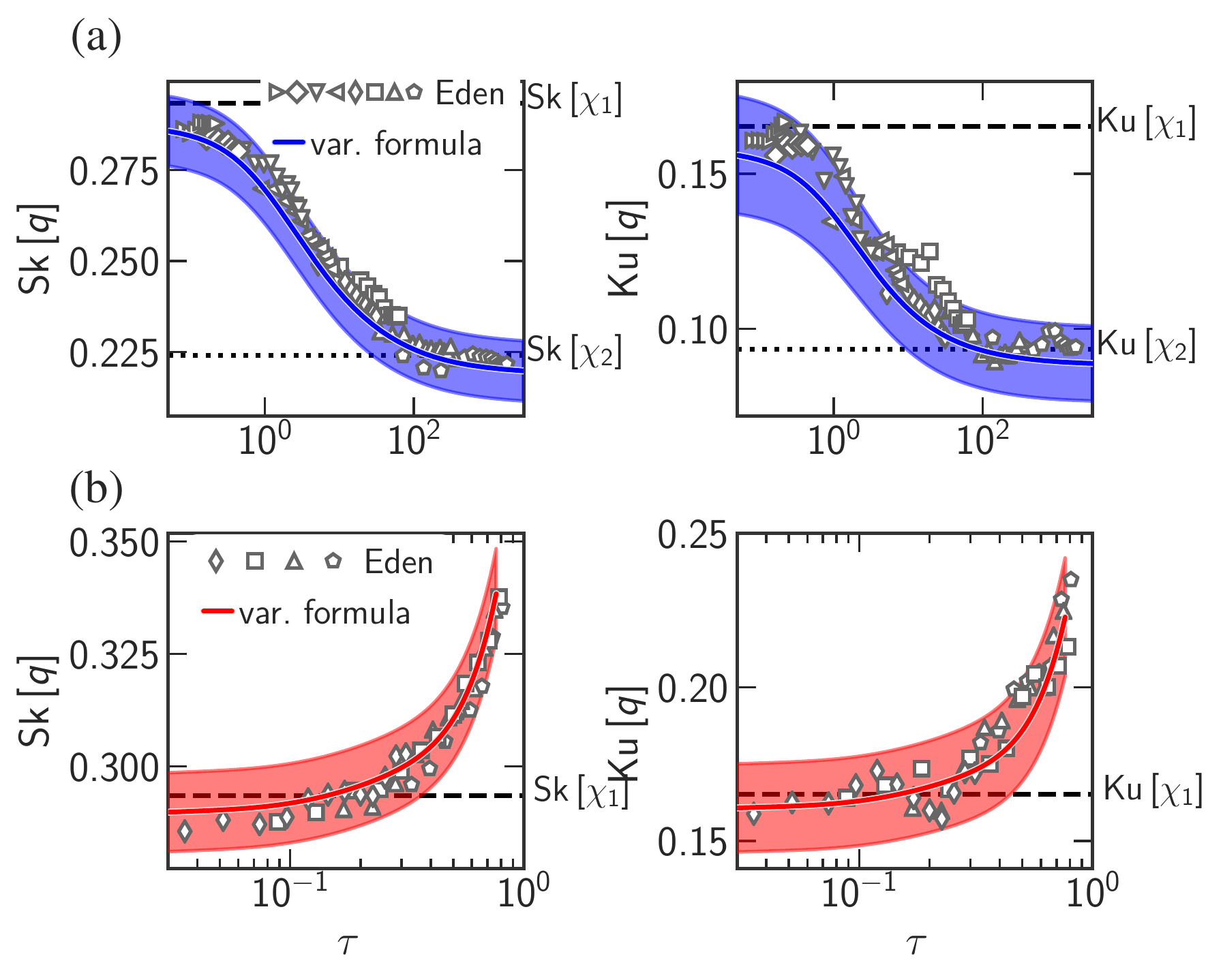}
	\caption{\label{fig:outin_skew_kurt}
		The skewness and kurtosis of the rescaled height $q(\theta,t)$, 
		plotted against the rescaled time $\tau=v_\infty{}t/R_0$
		for the (a) outgrowing and (b) ingrowing interfaces.
		The numerical results with $t>10^3$ are shown by gray open symbols,
		and the curves obtained by numerical evaluation of the variational (=var.) formula are drawn with the red solid line. 
		The symbols for the numerical results are for the samples with $N=40000(\rhd), 20000(\diamond), 10000(\triangledown), 4000(\lhd), 1000(\Diamond), \allowbreak 500(\square), 100 (\triangle), 50 (\pentagon)$ for the outgrowing interfaces, 
		and for $N=100000(\lhd), 40000(\Diamond), 20000(\square), 10000(\triangle)$ for the ingrowing interfaces, respectively. 
		The shaded area for the variational formula indicates uncertainty of the Monte-Carlo evaluation.
		The values for $\chi_1$ (flat) and $\chi_2$ (circular) are shown by the dashed and dotted lines, respectively. 
	}
\end{figure}

We evaluated the variational formula \eqref{eq:var_formula_parabola_general}
 by using Dyson's Brownian motion to approximate the Airy$_2$ process,
 as follows: 
\begin{enumerate}
\item 
We first prepared an initial Hermitian random matrix drawn from the Gaussian unitary ensemble (GUE), or equivalently, the stationary distribution of Eq.~\eqref{eq:OU_matrix} (below): 
\begin{equation}
H_{jk}(0)=
\begin{cases} 
\sqrt{\frac{1}{2}}\mathcal{N}^{(1)}_{jk} & (j=k) \\
\frac{1}{2}\left[\mathcal{N}^{(1)}_{jk} + \mathrm{i} \mathcal{N}^{(2)}_{jk} \right] & (j > k) 
\end{cases},
\end{equation}	
where $\mathcal{N}^{(m)}_{jk}$ are i.i.d.~random variables drawn from the normal distribution with the mean $0$ and the variance $1$.
\item 
We simulated the following Ornstein-Uhlenbeck process: 
\begin{equation}
\frac{\rd H_{jk}(u)}{\rd u} = 
\begin{cases} 
-H_{jk}+\eta^{(1)}_{jk}(u) & (j=k) \\
-H_{jk}+\sqrt{\frac{1}{2}}\left[\eta^{(1)}_{jk}(u) + \mathrm{i} \eta^{(2)}_{jk}(u)\right] & (j > k)
\label{eq:OU_matrix}
\end{cases}
\end{equation}
 with Gaussian noise $\eta^{(m)}_{jk}(u)$ satisfying 
$\left<\eta^{(m)}_{jk}(u)\right>\allowbreak=\allowbreak0$ and 
$\left<\eta^{(m')}_{j'k'}(u')\eta^{(m)}_{jk}(u)\right>
=\allowbreak\delta_{j'j}\allowbreak\delta_{k'k}\allowbreak\delta_{m'm}\allowbreak\delta(u'-u)$.
We employed Gillespie's exact algorithm for the Ornstein-Uhlenbeck process
 \cite{gillespie_1996} for each element, with time step $\Delta u$.
\item
We computed the largest eigenvalue $\lambda^{(N)}_j:=\lambda^{(N)}(j\Delta u)$ for each time step $j\Delta u$ $(j=0,\dots,j_\mathrm{max})$. 
Then, since
\begin{equation}
\sqrt{2}N^{1/6}\left[\lambda^{(N)}(N^{-1/3}Y)-\sqrt{2N}\right] \tod \mathcal{A}_2(Y) \qquad (N\to\infty),  \label{eq:DBM_Airy2}
\end{equation}
 we rescaled it as follows:
\begin{equation}
 \lambdarescaled{j}:=\sqrt{2}N^{1/6} \left(\lambda^{(N)}_j-\sqrt{2N}\right). \qquad
 \Delta\tilde{u}:= N^{1/3}\Delta u.
\end{equation}
\item 
Using the fact that the $\Airy{2}$ process is locally equivalent
 to the Brownian motion with unit diffusion constant
 (the standard Brownian motion) \cite{prahofer_2002,hagg_2008}, we approximated
 the variational formula \eqref{eq:var_formula_parabola_general} by 
\begin{equation}
\chiprbl{c} = \sup_{Y\in\mathbb{R}}\left[\Airyproc{2}{Y} -(1+c\tau)Y^2\right] \approx \max_{k=-L,\dots,L-1} r^{(j)}_{k} =:\chiprblN{c},  \label{eq:var_formula_ring}
\end{equation}
 where $L$ was taken sufficiently large (see below),
 $j=L,\dots,j_\mathrm{max}-L$,
 and $r^{(j)}_{k}$ is the maximum of the Brownian bridge
 (with unit diffusion constant) connecting 
$z^{(j)}_k:=\lambdarescaled{j+k} -(1+c\tau)\left(k\Delta \tilde{u}\right)^2$ 
and $z^{(j)}_{k+1}$.
Specifically, $r^{(j)}_{k}$ is a random variable whose cumulative distribution function is given by \cite{beghin_1999} 
\begin{equation}
	\mathbb{P}\left[r^{(j)}_{k} < z\right] = 
	\begin{cases}
		 1 - \exp\left[-\frac{(z-z^{(j)}_k)(z-z^{(j)}_{k+1})}{\Delta\tilde{u} } \right] & (z\ge\max\left\{z^{(j)}_k,z^{(j)}_{k+1}\right\}) \\
		 0 & (\text{otherwise}). 
	\end{cases}
\end{equation} 
The range of $L$ was chosen so that it satisfies
 $\left(\max\lambdarescaled{j} - \min\lambdarescaled{j}\right) < (1+c\tau) [(L/1.1) \Delta \tilde{u}]^2$.
Then the cumulants $\cum{\tilde{\chi}_N(c\tau)}{k}$ were evaluated by taking the average of the right-hand side of Eq.~\eqref{eq:var_formula_ring}
over varying $j$ and independent realizations of the Ornstein-Uhlenbeck process.
In our simulations for the main results presented in Figs.~\ref{fig:out_cum}, \ref{fig:in_cum}, and \ref{fig:outin_skew_kurt},
 we used $5360$ realizations of the Ornstein-Uhlenbeck process
 with $N=512$, $\Delta\tilde{u}=10^{-3}$ (see below),
 and $j_\mathrm{max}=500000$. 
\end{enumerate}

\subsection{Step size}
To find an appropriate step size $\Delta u$, 
 we estimated the range of $Y$ that is relevant
 to the value of the supremum in Eq.~\eqref{eq:var_formula_ring}. 
Since the Airy$_2$ process is locally equivalent to the standard Brownian motion, a value of $Y$ such that $(1+c\tau)Y^2$ becomes as large as $\Airyproc{2}{Y}$, denoted by $Y_0$, is given approximately by
$(1+c\tau) Y_0^2 = \sqrt{2Y_0}$.
With this $Y_0$, $\Delta\tilde{u}$ should be chosen so that $\Delta\tilde{u} \ll Y_0$.
For large $\tau$, $Y_0 \approx \tau^{-2/3}$.
In the present work,
 we chose $\Delta\tilde{u}=10^{-3}$ ($\Delta u = 10^{-3}N^{-1/3}$)
 so that $\Delta\tilde{u} \ll Y_0$ is satisfied for all $\tau\le3\times10^{3}$.

\subsection{Matrix size $N$}
\begin{figure}[H]
	\includegraphics[width=\linewidth]{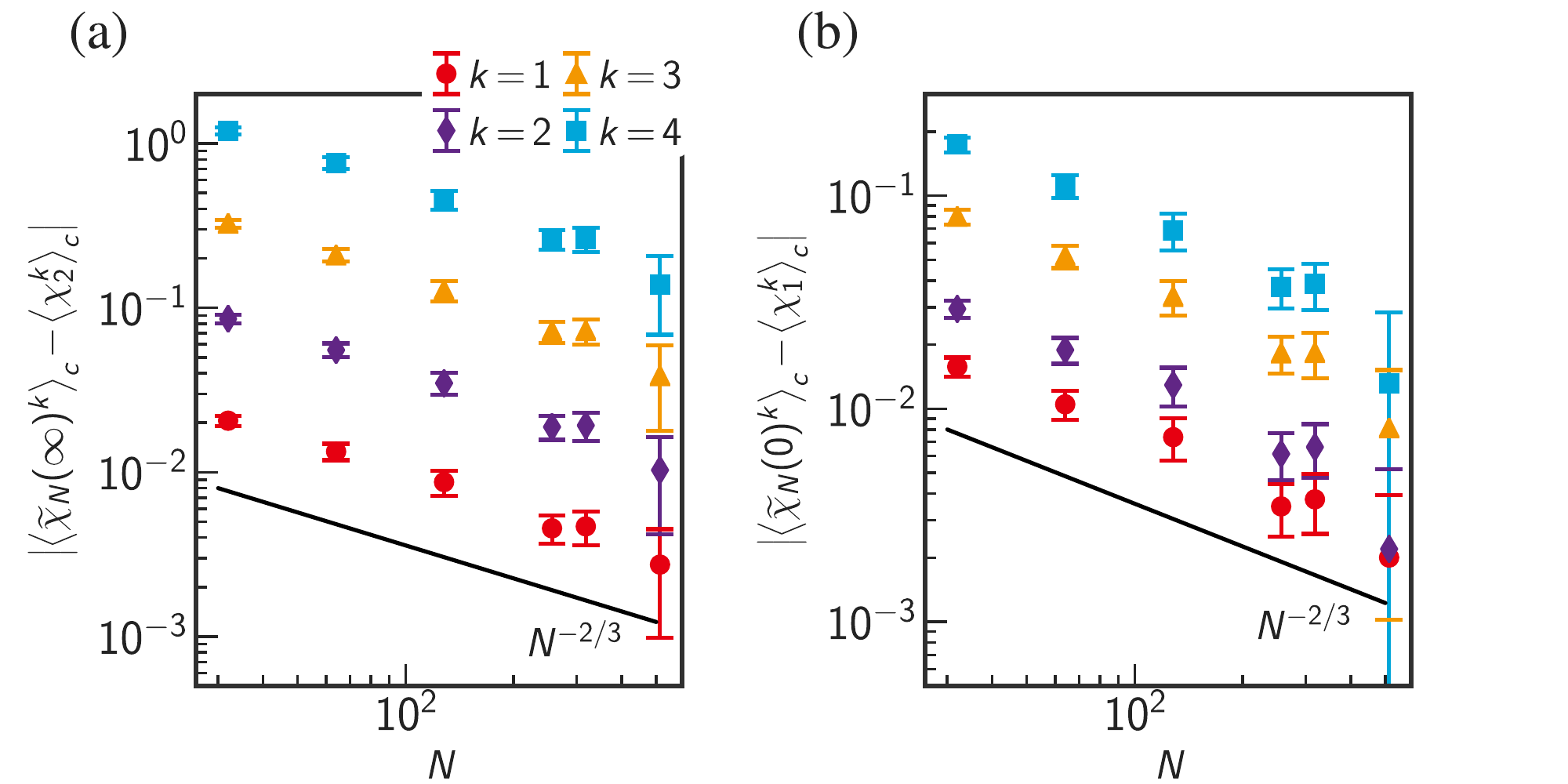}
	\caption{\label{fig:var_N_effect}
		Finite-$N$ corrections in the cumulants of $\tilde{\chi}_N(\infty) = \tilde{\lambda}_j^{(N)}$ (a) and $\tilde{\chi}_N\left(0\right)$ (b). They are compared with the cumulants of $\chi_2$ and $\chi_1$, respectively. 
		The errorbars indicate statistical uncertainty. The black solid lines are guides for the eyes showing the exponent $-2/3$.
	}
\end{figure}

To quantify the effect of finite matrix size $N$, 
 we first evaluated the cumulants of the rescaled largest eigenvalue,
 $\cum{(\tilde{\lambda}_j^{(N)})}{k}$, with varying $N$,
 and compared with the known values
 for the GUE Tracy-Widom distribution, $\cum{\chi_2}{k}$.
The results in Fig.~\ref{fig:var_N_effect}(a) show
 that $\cum{(\tilde{\lambda}_j^{(N)})}{k}$ indeed approach $\cum{\chi_2}{k}$,
 with the difference decreasing as $\sim N^{-2/3}$,
 being consistent with theoretical expectation \cite{johnstone_2012}.
Since $\tilde{\chi}_N(\infty) = \tilde{\lambda}_j^{(N)}$,
 the finite-$N$ corrections shown in Fig.~\ref{fig:var_N_effect}(a)
 are equivalent to those of $\tilde{\chi}(\tau)$ in the limit $\tau\to\infty$.
Similarly, we evaluated the variational formula \eqref{eq:var_formula_ring}
 with $c\tau=0$, for which it is known that $\tilde{\chi}(0)=\sup_{Y\in\mathbb{R}}(\Airyproc{2}{Y}-Y^2)\eqd\chi_1$,
 i.e., GOE Tracy-Widom distribution \cite{johansson_2003,Corwin.etal-CMP2013}.
As displayed in Fig.~\ref{fig:var_N_effect}(b), the data obtained
 from the variational formula indeed show the cumulant values
 approaching those of $\chi_1$, again with the finite-size corrections
 proportional to $N^{-2/3}$.
From those results, we decided to use $N=512$ (largest size)
 for the main part of the work;
 at $N=512$, the amplitude of the finite-$N$ corrections shown
 in Fig.~\ref{fig:var_N_effect} is small enough to compare with
 our numerical and experimental data.
The uncertainty of the estimated cumulants displayed
 in Figs.~\ref{fig:out_cum}, \ref{fig:in_cum}, and \ref{fig:outin_skew_kurt}
 is the summation of the statistical uncertainty
 and the expected amplitude of finite-$N$ correction,
 the latter being evaluated by the larger value of $\left|\cum{\tilde{\chi}_N\left(\infty\right)}{k}-\left<\chi_2^k\right>\right|$ and $\left|\cum{\tilde{\chi}_N\left(0\right)}{k}-\left<\chi_1^k\right>\right|$ [Fig.~\ref{fig:var_N_effect}(a) and (b), respectively].

\newpage
\twocolumngrid

\section{Supplemental Tables}

\begin{table}[H]
	\caption{\label{tab:simulation_params_eden}
		Parameters for the Eden simulations.}
	\centering\begin{tabular}[t]{|l|cccccc|} 
		\hline
		type & \multicolumn{6}{c|}{outgrowing} \\ 
		\hline
		$N$ & $50$ & $100$ & $500$ & $1000$ & $4000$ & $10000$ \\
		\# of samples & $14400$ & $14400$ & $14400$ & $14400$ & $11520$ & $4320$  \\
		\hline
		type & \multicolumn{2}{c|}{outgrowing} & \multicolumn{4}{c|}{ingrowing} \\ 
		\hline
		$N$ & $20000$ &  \multicolumn{1}{c|}{$40000$} & $10000$ & $20000$ & $40000$ &  \multicolumn{1}{c|}{$100000$}  \\
		\# of samples & $4320$ &  \multicolumn{1}{c|}{$4320$} & $4320$ & $4320$ & $4320$ & \multicolumn{1}{c|}{$1440$}  \\
		\hline
type & \multicolumn{1}{c|}{flat} \\ 
\cline{1-2}
length & \multicolumn{1}{c|}{$75000$}  \\
\# of samples & \multicolumn{1}{c|}{$14400$}  \\
\cline{1-2}
	\end{tabular}
\end{table}

\begin{table}[H]
	\caption{\label{tab:nonuniversal_params_lc}
		Experimental conditions and non-universal parameters. }
	\centering\begin{threeparttable}
		\begin{tabular}{|l|c|c|c|c|} \hline
			initial condition
			& \# of samples
			& $v_\infty (\si{\um \per s})$
			& $\Gamma (\si{\um^3 \per s})$
			\\
			\hline
			line   & $1417$& $30.84(2)$ & $\SI{1.25(2)e3}{}$ \\
			point & $941$ & $29.68(3)$ & $\SI{1.31(4)e3}{}$ \\
			$R_0=\SI{366}{\um}$  & $1936$ & $30.84(2)$\tnote{*} & -- \\
			$R_0=\SI{219}{\um}$  & $1521$ & $30.60(2)$\tnote{*} & -- \\ \hline
		\end{tabular}
		\begin{tablenotes}[normal]
			\item[*] Values obtained by fitting the last data point of $\left<p\right>$ to the results of the Eden model (see main text).
		\end{tablenotes}
	\end{threeparttable}
\end{table}

\section{Supplemental Figures}

\begin{figure}[H]
	\includegraphics[width=\linewidth]{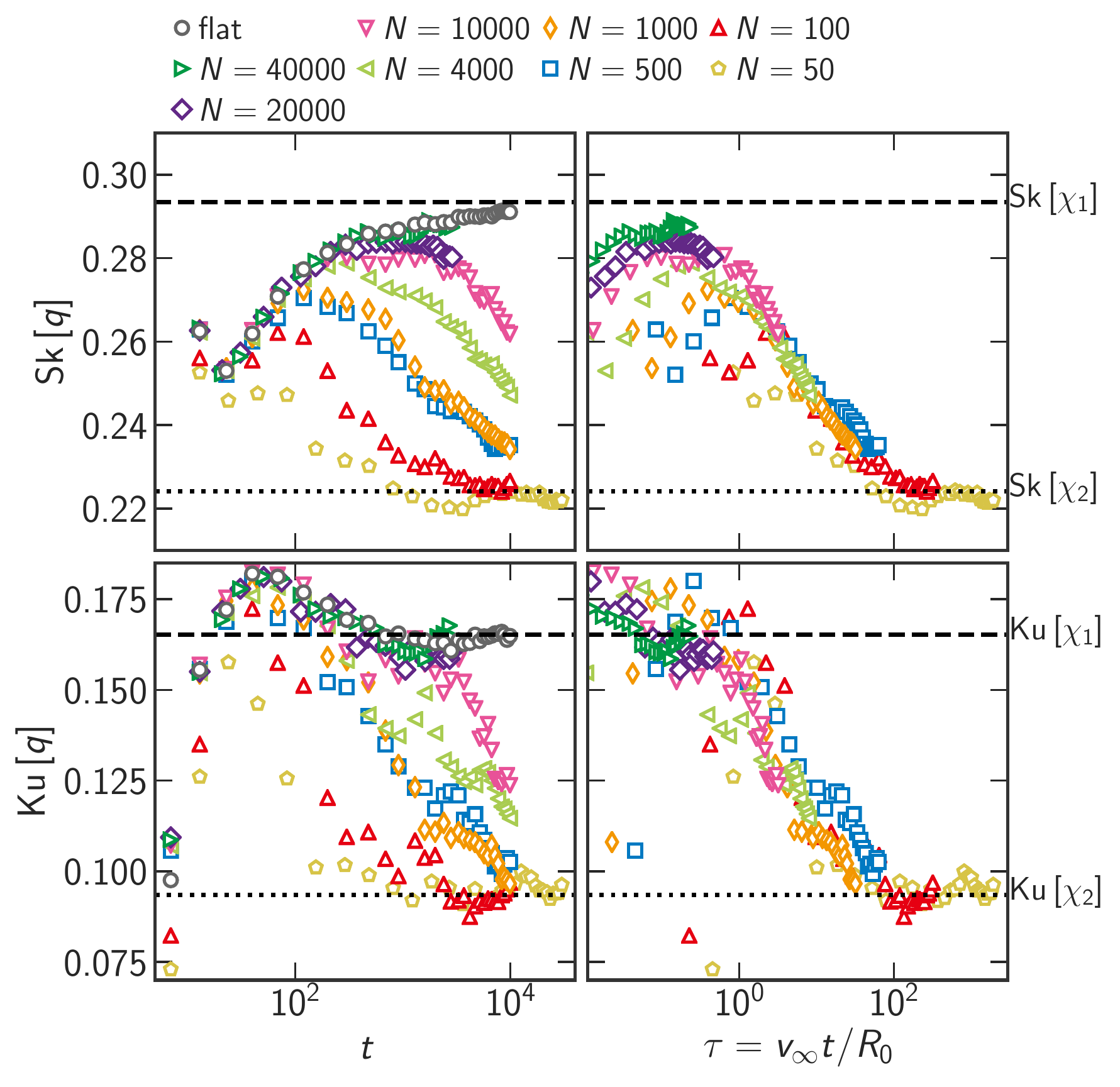}
	\caption{\label{fig:eden_skew_kurt}
		The skewness and kurtosis of the rescaled height $q(\theta,t)$ for the Eden model. The data are plotted against the raw time $t$ (left column) and the rescaled time $\tau=v_\infty t/R_0$ (right column). The values for $\chi_1$ and $\chi_2$ are shown by the dashed and dotted lines, respectively. }
\end{figure}

\begin{figure}[H]
	\centering
	\includegraphics[width=0.8\linewidth]{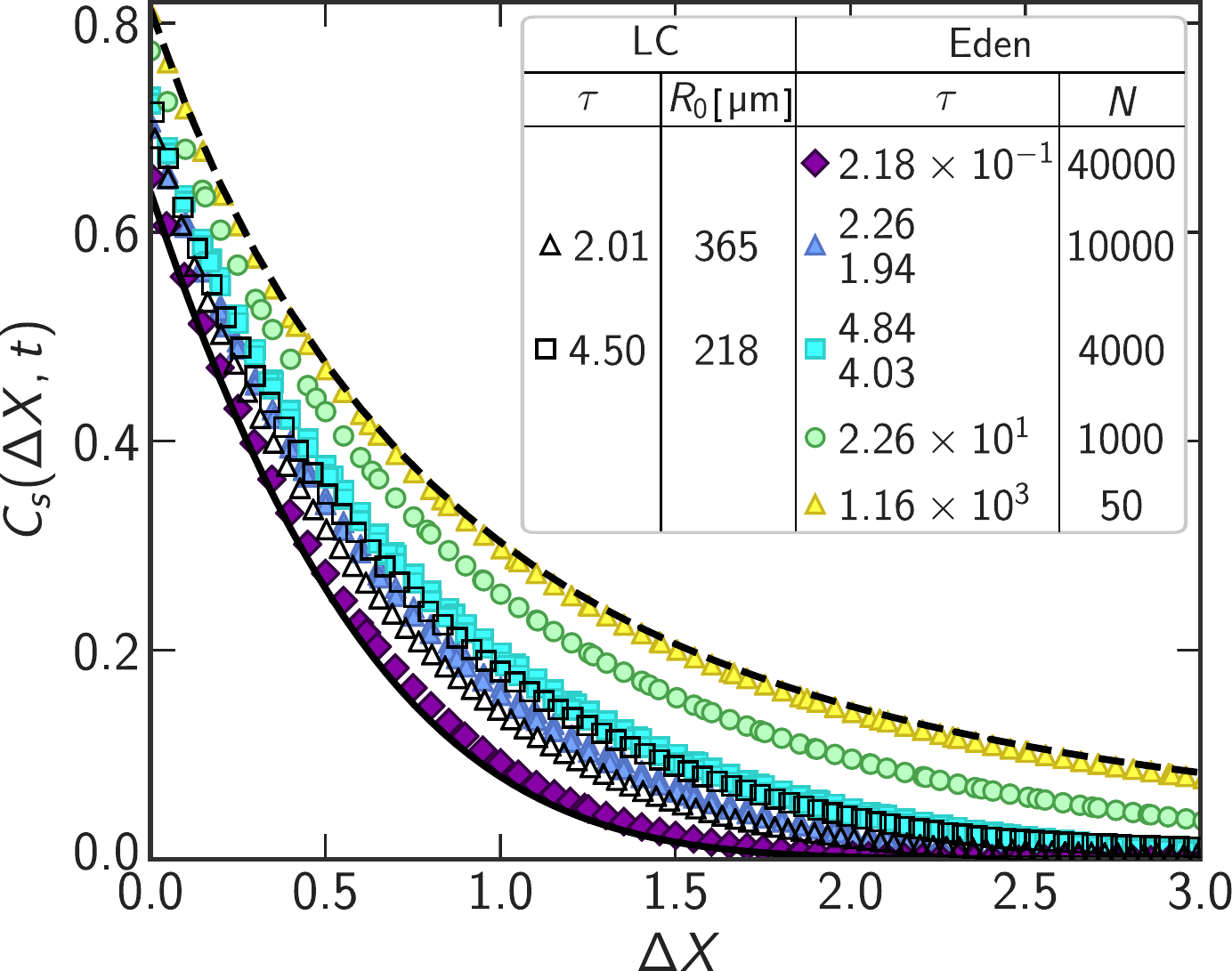}
	\caption{\label{fig:spatial_corr}
		Spatial covariance $C_s(\Delta X,t)$, plotted against the normalized length $\Delta X$,
		for the outgrowing interfaces.
		The solid and dashed lines indicate
		the $\text{Airy}_1$ (flat) and $\text{Airy}_2$ (circular) covariance, respectively, $\expct{\mathcal{A}_i(X+\Delta{}X)\mathcal{A}_i(X)} - \expct{\mathcal{A}_i(X)}^2$.
	}
\end{figure}

\bibliography{citations}